\documentclass[aps,preprint]{revtex4}
\usepackage{amsfonts}
\usepackage{amsmath}
\usepackage{amssymb}
\usepackage{graphicx}

\input{tcilatex}

\begin{document}

\title{Condensation and evolution of space-time network}
\author{Bi Qiao$^{1,2}$}
\affiliation{$^{1}$Department of Physics, Science School, Wuhan University of Technology,
Wuhan 430070, China; $^{2}$International Noble Academy, 1075 Ellesmere Road,
Toronto, M1P 5C3 Canada}

\begin{abstract}
In this work, we try to propose, in a novel way using the Bose and Fermi
quantum network approach, a framework studying condensation and evolution of
space time network described by the Loop quantum gravity. Considering
quantum network connectivity features in the Loop quantum gravity, we
introduce a link operator, and through extending the dynamical equation for
the evolution of quantum network posed by Ginestra Bianconi to an operator
equation, we get the solution of the link operator. This solution is
relevant to the Hamiltonian of the network, and then is related to the
energy distribution of network nodes. Showing that tremendous energy
distribution induce huge curved space-time network, may have space time
condensation in high-energy nodes. For example, in the black hole
circumstances, quantum energy distribution is related to the area, thus the
eigenvalues of the link operator of the nodes can be related to quantum
number of area, and the eigenvectors are just the spin network states. This
reveals that the degree distribution of nodes for space-time network is
quantized, which can form the space-time network condensation. The black
hole is a sort of result of space-time network condensation, however there
may be more extensive space-time network condensation, for example, the
universe singularity (big bang).

\textbf{Keywords: }Loop quantum gravity, Spin network, Complex network,
Quantum network, Black hole.
\end{abstract}

\startpage{0}
\endpage{20}
\maketitle

\section{Introduction}

The greatest challenging topic to contemporary fundamental physics is to
coordinate general relativity and quantum mechanics, quantum gravity
research. This study will cause us to have profound changes in the concept
on space, time and matter structure, and achieve a quantum revolution. At
present, for the quite progressed theory of quantum gravity, there are two
forms. First, quantized general relativity, canonical quantum gravity
belongs to this. Second, quantized a classic theory which is deferent from
general relativity and the general relativity is as its low-energy limit.
The super-string / M theory belongs to this. Loop quantum gravity $^{\text{%
\cite{1}}}$ is the current representative form of canonical quantum gravity.
Canonical quantum gravity is one only when the gravitational effects of
quantum gravity. Comparing with superstring / M theory it does not include
other different interactions. Its basic concept is to apply standard
quantization procedure to general relativity while general relativity is
written in the canonical (Hamiltonian) form. According to historical
development the canonical quantum gravity can be divided into two stages,
the simple quantum gravity and Loop quantum gravity. Roughly speaking, the
former occurred before 1986, and later occurred after 1986. In simple
quantum gravity, there is renormalization difficulty for UV divergence, and
Loop quantum gravity has become currently the representative.

According to basic spirit of the Loop quantum gravity space-time is the
gravitational field and space-time is quantized in the Planck scale, as well
formed space-time network which uses quantum of the volume as a node,
connected through the quantum of adjacent area. The basis of the space time
network can be described by spin network. Physical space time can be
expanded by spin networks as basis. The evolution of spin network can be
studied by calculating spinfoam. Spinfoam is main method to study evolution
of spin network, based on the Feynman propagation in the Loop quantum
gravity. The current studied complex network theoretical method $^{\text{%
\cite{2}}}$ is irrelevant to the study of the space time network.

In this work, we try to use the Bose and Fermi quantum network approach to
present a framework for studying condensation and evolution of space time
network. Considering quantum network connectivity features, we introduce a
link operator, and through extending the complex network equation for the
evolution of quantum network in the literature \cite{3} to an operator
equation, we get the solution of the link operator.

This solution is relevant to the Hamiltonian of the network, and then is
related to the energy distribution of network nodes. Showing that tremendous
energy distribution induce huge curved space-time network, may have space
time condensation in high-energy nodes. For example, in the black hole
circumstances, quantum energy distribution is related to the area, thus the
eigenvalues of the degree operator of the nodes can be related to quantum
number of area, and the eigenvectors are just the spin network states. This
reveals that the degree distribution of nodes for space-time network is
quantized, which can form the space-time network condensation. The black
hole is a sort of space-time network condensation, however there may be more
extensive network of space-time setting, for example, the universe
singularity (big bang).

\section{Connectivity of Spin Network}

Let a graph $\Gamma$ to be given immersed in the manifold $\Sigma$, for
which an ordering and an orientation have been chosen. Denote the nodes as
the end points of the oriented curves in $\Gamma$, and joined by links $l$.
Then $j_{l}$ is an assignment of an irreducible representation to each link $%
l$ and $i_{n}$ is an assignment of an intertwiner to each node $n$. The
intertwiner $i_{n}$ associated with a node is between the representations
associated with the links adjacent to the node. $\Gamma$ becomes a
collection of nodes $n$ with links $l$. The triplet $S\equiv\left(
\Gamma,i_{n},j_{n}\right) $ is called a spin network embedded in $\Sigma$.
The physical space-time is combination of spin networks. More precisely, $%
j_{n}$ represents unitary irreducible representations and $i_{n}$ represents
a basis in the space of the intwiners between the representations adjacent
to the nodes $n$. The area of a surface cutting $n$ links of spin network
with $l=1,\cdots,n$ quantized labels $j_{l}$, given by the eigenvalue of the
area operator, $8\pi\gamma \allowbreak\frac{h}{2\pi}G\sum_{n}\sqrt{%
j_{n}\left( j_{n}+1\right) }$, where $\gamma$ is the Immirzi parameter which
is a free dimensionless constant and $G$ is the Newton constant. Consider
the quantum property of the connectivity for nodes of spin network, we
introduce a link (degree) operator of node $\mathbf{K}$ which acts to the
spin network state $\left\vert S\right\rangle $ and gives the eigenvalue $%
j_{n}$ by 
\begin{equation}
\mathbf{K}\left\vert S\right\rangle =\tau^{f\left( j_{n}\right) }\left\vert
S\right\rangle ,   \label{1}
\end{equation}
where $f$ is a function of $j_{n}$, and $\left\vert S\right\rangle =\left(
\Gamma,i_{n},\alpha_{n},\beta_{n}\right) $ is given by the spin network $%
S\equiv\left( \Gamma,i_{n},j_{n}\right) $ in such way : $\left\langle
U\right\vert \left. j,\alpha,\beta\right\rangle =\left( R^{j}\left( U\right)
\right) _{\alpha}^{\beta}$ forms an orthonormal basis in the Hilbert space $%
L_{2}\left[ G,dU\right] $), and introduces a spin network state $\left\vert
S\right\rangle $ by the Peter-Weyl theorem$^{\text{\cite{1}}}$.

Extending the rate equation of connectivity in ref.[2] to the link operator $%
\mathbf{K}$, a dynamical equation for $\mathbf{K}$ can be constructed by 
\begin{equation}
\frac{\partial\mathbf{K}}{\partial\tau}=m_{+}\frac{e^{\beta H_{0}}\mathbf{K}%
}{Tr\left( e^{\beta H_{0}}\mathbf{K}\right) }-m_{-}\frac{e^{-\beta H_{0}}%
\mathbf{K}}{Tr\left( e^{-\beta H_{0}}\mathbf{K}\right) },   \label{2}
\end{equation}
where $\tau$ is introduced as a proper time in this relativity system, the
spin network state is associated with a general three-dimensional spacelike
hypersurface, $m_{+}$ are $m_{-}$ is a proportional (suitable) coefficients
(constant or function) for making the equation (\ref{2}) to be held, and $%
H_{0}$ is an energy part of the (constrained) Hamiltonian of the system $H$
given by 
\begin{align}
H & =\int Ntr\left( F\wedge\left\{ \mathbf{V},A\right\} \right)  \label{3} \\
& =p_{t}+H_{0},
\end{align}
here $F$ (curvature) and $A$ (complex field) are limits of holonomy
operators of small paths, $\left\{ {}\right\} $ is the Poisson bracket as
quantum commutator, and $\mathbf{V}$ is the quantum volume operator$^{\text{%
\cite{1},\cite{1b}}}$.

In the thermodynamical limit, $Tr\left( e^{\beta H_{0}}\mathbf{K}\right) $
and $Tr\left( e^{-\beta H_{0}}\mathbf{K}\right) $ are supposed to tend to%
\begin{align}
Tr\left( e^{\beta H_{0}}\mathbf{K}\right) & \rightarrow m_{+}\tau
e^{\beta\mu_{+}},  \label{4} \\
Tr\left( e^{-\beta H_{0}}\mathbf{K}\right) & \rightarrow m_{-}\tau
e^{-\beta\mu_{-}},   \label{5}
\end{align}
where $\mu_{+}$, $\mu_{-}$ are suitable constants which are related to the
asymptotic behavior of the network. This is so because $Tr\left( e^{\beta
H_{0}}\mathbf{K}\right) $ and $Tr\left( e^{-\beta H_{0}}\mathbf{K}\right) $
can be assumed to be self-average and converge to their mean value and grow
linearly in time under suitable conditions$^{\text{\cite{4}}}$.

Thus, Eq.(\ref{2}) becomes 
\begin{equation}
\frac{\partial\mathbf{K}}{\partial\tau}=\left[ e^{\beta\left( H_{0}-\mu
_{+}\right) }-e^{-\beta\left( H_{0}-\mu_{-}\right) }\right] \frac {\mathbf{K}%
}{\tau},   \label{6}
\end{equation}
and the solution of the equation therefore is given by%
\begin{equation}
\mathbf{K}=\tau^{f\left( H_{0}\right) },   \label{7}
\end{equation}
where $f\left( H_{0}\right) $ is defined by 
\begin{equation}
f\left( H_{0}\right) =e^{\beta\left( H_{0}-\mu_{+}\right) }-e^{-\beta \left(
H_{0}-\mu_{-}\right) }.   \label{8}
\end{equation}
This allows one to obtain the mean value $\left\langle \mathbf{K}%
\right\rangle $ as%
\begin{equation}
\left\langle \mathbf{K}\right\rangle =\left( m_{+}+m_{-}\right) \left( \frac{%
\tau}{\tau^{\prime}}\right) ^{f\left( M\right) },   \label{9}
\end{equation}
with the initial condition%
\begin{equation}
\left\langle \mathbf{K}\right\rangle _{0}=m_{+}+m_{-},   \label{9a}
\end{equation}
where $M$ represents the eigenvalue of $H_{0}$ which is related to the
energy in the nodes, and relevant tracing calculation is performed with
respect to the eigen-states of $H_{0}$. When $H_{0}$ acts on a spin network
state, this operator acts only on the nodes of the spin network, because of
the presence of the volume in the Hamiltonian operator $H$, hence $%
\left\langle \mathbf{K}\right\rangle $ can represent a general connectivity
of space-time network, this space-time network can be expanded by many spin
networks as unit networks.

\section{Condensation of space-time network}

In the Schwarzschild black hole environment, the energy $M$ is related to
the area $\emph{A}$ through$^{\text{\cite{5}}}$ 
\begin{equation}
M=\sqrt{\frac{\emph{A}}{16\pi G^{2}}}.   \label{10}
\end{equation}
Thus, the mean value $\left\langle \mathbf{K}\right\rangle $ becomes%
\begin{equation}
\left\langle S\right\vert \mathbf{K}\left\vert S\right\rangle =\left(
m_{+}+m_{-}\right) \left( \frac{\tau}{\tau^{\prime}}\right) ^{f\left( \sqrt{%
\frac{\emph{A}}{16\pi G^{2}}}\right) }=\left( m_{+}+m_{-}\right) \left( 
\frac{\tau}{\tau^{\prime}}\right) ^{f\left( \sqrt{\frac{\frac {h}{2\pi}\gamma%
\sqrt{j_{n}\left( j_{n}+1\right) }}{2G}}\right) },   \label{11}
\end{equation}
where $f\left( \sqrt{\frac{1}{2G}\frac{h}{2\pi}\gamma\sqrt{j_{n}\left(
j_{n}+1\right) }}\right) $ is a function of $\sqrt{j_{n}\left(
j_{n}+1\right) }$ or $j_{n}$. This shows that $\left\langle S\right\vert 
\mathbf{K}\left\vert S\right\rangle $ is quantized, with quantum number of
link as $\sqrt{j_{n}\left( j_{n}+1\right) }$, where $j_{n}$ is determined by
eigenvalues of the area operator $\mathbf{A}$ with respect to spin network
state $S$ and represents the quantum number of the area. One of properties
of the quantized links of nodes is that there are possibly many links
labelled by quantum numbers between two adjacent nodes with relevant quantum
weights, while the classical link between the same two nodes is just one
link with (or without) a classical weight.

Consequently, one can get%
\begin{equation}
\frac{m}{m+m^{\prime}}=\int dMp\left( M\right) \frac{e^{\beta\left(
M-\mu_{+}\right) }}{1-e^{\beta\left( M-\mu_{+}\right) }+e^{-\beta\left(
M-\mu_{-}\right) }},   \label{12}
\end{equation}
and%
\begin{equation}
\frac{m^{\prime}}{m+m^{\prime}}=\int dMp\left( M\right) \frac{e^{-\beta
\left( M-\mu_{+}\right) }}{1-e^{\beta\left( M-\mu_{+}\right)
}+e^{-\beta\left( M-\mu_{-}\right) }}.   \label{13}
\end{equation}
Here the ratio $\frac{m}{m+m^{\prime}}$ or $\frac{m^{\prime}}{m+m^{\prime}}$
represents a fraction of $m$ or $m^{\prime}$ links attached or detached to
nodes in the evolution of the network. The former represents the increase of
the links on the node, and the second one represents the decrease of links
on the node. When $\mu_{+}=\mu_{-}=0$ and $M\longrightarrow0$, it is%
\begin{equation}
\frac{m}{m+m^{\prime}}=\frac{m^{\prime}}{m+m^{\prime}}=\int dMp\left(
M\right) ,   \label{15}
\end{equation}
which shows that the quantized connectivity of the space-time network, in
the very low energy condition, evolves either increase or decrease in time
as a mixed process of attached or detached to nodes. While $\mu_{+}=\mu_{-}=0
$ and $M\longrightarrow\infty$, it is not difficult to find $\frac{m^{\prime}%
}{m+m^{\prime}}=0$, and one gets 
\begin{equation}
\frac{m}{m+m^{\prime}}=-\int dMp\left( M\right) .   \label{14}
\end{equation}
This shows, in this situation, there is only link increased process on the
node while the detached process is vanished! This reveals that a kind of
condensation of space-time network around the huge energy nodes may happen
because there is only a monotonic attached link process in the evolution of
the space-time network. For example, in the (Schwarzschild) black hole
surroundings, there is a sort of space-time network condensation on the
singularity with $M\longrightarrow\infty$. Furthermore, in the universal
space-time scale, there may be more extensive space-time network
condensation, such as the universe condensates itself as one (many)
singularity, a (many) big bang.

\section{Connectivity distribution}

The connectivity distribution $P\left( \left\langle \mathbf{k}\right\rangle
\right) $ is given by the sum of the probabilities of the nodes with energy $%
M$ and connectivity $\left\langle \mathbf{k}\right\rangle $. Following the
idea in ref.\cite{2}, considering the degree of the nodes in our model $k$
changes to the mean value $\left\langle \mathbf{K}\right\rangle $ since the
quantum property of the link for the space-time network, we have to sum over
all the modes with energy lower or higher than a threshold defined as $%
\epsilon_{s}=\frac{\mu_{+}+\mu_{-}}{2}$, when $\left\langle \mathbf{k}%
\right\rangle <m+m^{\prime}$ or $\left\langle \mathbf{k}\right\rangle
>m+m^{\prime}$, respectively,%
\begin{align}
P\left( \left\langle \mathbf{k}\right\rangle \right) & =\theta\left(
\left\langle \mathbf{k}\right\rangle -\left\langle \mathbf{k}\right\rangle
_{0}\right) \int_{\epsilon_{s}>\frac{\mu_{+}+\mu_{-}}{2}}dMp\left( M\right) 
\frac{1}{\frac{\partial\left\langle \mathbf{k}\right\rangle }{\partial\tau}}%
+\theta\left( \left\langle \mathbf{k}\right\rangle _{0}-\left\langle \mathbf{%
k}\right\rangle \right) \int_{\epsilon_{s}<\frac{\mu_{+}+\mu_{-}}{2}%
}dMp\left( M\right) \frac{1}{\frac{\partial\left\langle \mathbf{k}%
\right\rangle }{\partial\tau}}  \label{b2} \\
& =\theta\left( \left\langle \mathbf{k}\right\rangle -m-m^{\prime}\right) 
\frac{1}{m+m^{\prime}}\int_{\epsilon_{s}>\frac{\mu_{+}+\mu_{-}}{2}}dMp\left(
M\right) \frac{1}{f\left( M\right) }\left( \frac{\left\langle \mathbf{k}%
\right\rangle }{m+m^{\prime}}\right) ^{-1-\frac{1}{f\left( M\right) }} 
\notag \\
& +\theta\left( m+m^{\prime}-\left\langle \mathbf{k}\right\rangle \right) 
\frac{1}{m+m^{\prime}}\int_{\epsilon_{s}<\frac{\mu_{+}+\mu_{-}}{2}}dMp\left(
M\right) \frac{1}{f\left( M\right) }\left( \frac{\left\langle \mathbf{k}%
\right\rangle }{m+m^{\prime}}\right) ^{-1-\frac{1}{f\left( M\right) }}. 
\notag
\end{align}
If $\mu_{+}=\mu_{-}=0$ and $M\longrightarrow\infty$, it is possible that $%
\frac{1}{f\left( M\right) }=0$, and $\int_{\epsilon_{s}>\frac{\mu_{+}+\mu_{-}%
}{2}}dMp\left( M\right) \frac{1}{f\left( M\right) }$ is finite (convergent
and $\neq0$), which results in%
\begin{equation}
P\left( \left\langle \mathbf{k}\right\rangle \right) =\theta\left(
\left\langle \mathbf{k}\right\rangle -m-m^{\prime}\right) \frac {1}{%
m+m^{\prime}}\int_{\epsilon_{s}>\frac{\mu_{+}+\mu_{-}}{2}}dMp\left( M\right) 
\frac{1}{f\left( M\right) }\left( \frac{\left\langle \mathbf{k}\right\rangle 
}{m+m^{\prime}}\right) ^{-1}\sim B\frac {1}{\left\langle \mathbf{k}%
\right\rangle },   \label{bb}
\end{equation}
where the coefficient $B$ is defined by%
\begin{equation}
B=\int_{\epsilon_{s}>\frac{\mu_{+}+\mu_{-}}{2}}dMp\left( M\right) \frac {1}{%
f\left( M\right) },   \label{b3}
\end{equation}
and the integral variable is supposed to be irrelevant to $\left\langle 
\mathbf{k}\right\rangle $. This result means that the connectivity
distribution $P\left( \left\langle \mathbf{k}\right\rangle \right) ,$ in the
above approximations, obeys the power law and the space-time network is
scale-free, the connectivity distribution $P\left( \left\langle \mathbf{k}%
\right\rangle \right) $ is inversely proportional to the increase of the
connectivity $\left\langle \mathbf{k}\right\rangle $. This means the
probability of $\left\langle \mathbf{k}\right\rangle $ tending to infinity
is quite small if value of $B$ is tiny. However, if $B$ is big or is a
function of $\left\langle \mathbf{k}\right\rangle $, then situation of $%
P\left( \left\langle \mathbf{k}\right\rangle \right) $ is complicated, then,
the relevant space-time network may be not scale-free.

\section{Conclusions}

The quantum connectivity of space-time network can be calculated by
introducing the link operator. The mean value of the link operator is a
function of eigenvalue for the energy part of the Hamiltonian operator of
the space-time network. In the Schwarzschild black hole situation, the mean
value of the link operator is a function of quantum number of the area
operator with respect to the spin network states, and the connectivity is
quantized. Moreover, there is a sort of condensation of space-time network
around its extreme high energy nodes. In the condensation progression there
exists only a monotonic attached link process in the evolution of the
network, which may results in the singularity in the universe. The
connectivity distribution can be represented by Eq.(\ref{bb}), which may or
may not obey the power law.

\textbf{Acknowledgement: }This work was supported by the grants from NSERC,
CIPI, MMO, CITO, Wuhan University of Technology, and Chinese NSFC under the
Grand No. 60874087.

\end{document}